\documentclass{elsart}

\usepackage{graphicx}

\usepackage{amssymb}

\begin{document}

\begin{frontmatter}

\title{Influence of hydrodynamics on the interpretation of the high
$p_{T}$ hadron suppression at RHIC}

\author{T.~Peitzmann\thanksref{email}}

\address{Utrecht University, NL-3508 TA Utrecht, The Netherlands}

\thanks[email]{email: t.peitzmann@phys.uu.nl}
\begin{abstract}
A hybrid parameterization including contributions of hydrodynamics and 
of expectations from the spectral shape observed in p+p collisions is
introduced. This parameterization can successfully describe
identified hadron spectra over a wide range of $p_{T}$ in Au+Au
reactions at $\sqrt{s_{_{NN}}}$ = 200~GeV for all centralities. The
parameters of the hydrodynamic source compare reasonably well to
other attempts to describe the spectra. The description is obtained
using one universal suppression
factor of the hard scattering component independent of $p_{T}$ and
hadron species. For the fit results obtained the observed nuclear
modification factor for the different particles
converges to a universal suppression behavior for $p_{T} > 6 \,
\mathrm{GeV}/c$.
\end{abstract}

\end{frontmatter}

\section{Introduction}
\label{sec:intro}
One of the most interesting recent observations in ultrarelativistic 
heavy-ion physics was the suppressed 
yield of moderately high $p_T$ neutral pions in central
Au+Au collisions at $\sqrt{s_{_{NN}}}$ = 130~GeV with respect 
to appropriately scaled p+p results~\cite{phenix:raa}, in contrast to
a strong enhancement observed at lower beam energies~\cite{wa98}. This 
was qualitatively supported by the observation of a suppression in
non-identified charged hadron yields~\cite{phenix:raa,star:raa}. 
Among the theoretical studies on the origin of the observed high $p_T$ 
deficit~\cite{vitev,ina,levai,xnwang,dima,gallmeister}, most are
based on the early prediction~\cite{Gyu90} of the so-called jet
quenching, i.e. the energy loss of fast partons induced by a hot and 
dense quark-gluon-plasma. Others invoke initial state parton
saturation~\cite{dima} or final state hadronic
interactions~\cite{gallmeister}. While these different scenarios are 
being discussed, the interpretation is complicated by the parallel
observation of a comparably large proton and antiproton yield at intermediate
momenta~\cite{phenix:hadrons}, which are apparently not similarly
suppressed. These observations have been confirmed by the new
measurements at $\sqrt{s_{_{NN}}}$ =
200~GeV~\cite{qm02:saskia,qm02:kunde}, which extend out
to considerably higher transverse momenta for the pion measurement.

Simultaneously, there are a number of hints from the RHIC
experiments that equilibration may be achieved already in an early
state and that the behavior of the system might at least partly be described by
hydrodynamics, most notably the observation of strong elliptic
flow~\cite{star:flow,phenix:flow}. Also the hadron momentum spectra
can be described by hydrodynamic calculations and parameterizations
quite well in the low and intermediate momentum 
range~\cite{tp:hydro,broniowski,qm02:jane,hydro:kolb}. In particular, the large 
(anti)proton/pion ratio can easily be explained by such calculations.
While at very high $p_{T}$ the influence of hydrodynamic production
should become negligible, in the intermediate momentum range there
will be a smooth transition from hydrodynamic behavior to hard
scattering. 
It is therefore of interest to study the implications of hydrodynamic
particle production for the interpretation of the pion suppression.  

In the present paper I will attempt to describe hadron spectra with a
para\-me\-terization combining hydrodynamic components and components
similar to the original particle production in p+p collisions. This
is considered a simple approximation to the more complicated real
situation in heavy ion collisions, where, even if hydrodynamics is
relevant for a large fraction of the particle spectra, there will
always be some non-equilibrium contribution e.g. from the surface of 
the collision zone. In reality, part of this would likely be
intermediate in shape between true p+p spectra and hydrodynamics, but
it is beyond the scope of this paper to attempt to describe the full 
non-equilibrium nature of these collisions. The calculations should
nevertheless provide a better understanding of the hadron spectra
than pure hydrodynamic calculations. Furthermore, the inclusion of an 
explicit hard component will allow to describe the spectra out to
high $p_{T}$.

\section{The model}
\label{sec:model}

\subsection{The hydrodynamical parameterization}
\label{sec:hydro}
The present paper uses an extension of the hydrodynamical
parameterization of \cite{tp:hydro}, which includes effects of 
transverse flow and resonance decays. It was originally based on a
computer program by Wiedemann and Heinz \cite{wiedemann96}, which in 
turn builds upon \cite{Schn93}. It
parameterizes the particle source at freeze-out as:
\begin{eqnarray}
\label{eq:urs1}
  {dN_r^{\rm dir}\over dY dM_T^2} 
  &=& N_{hydro} \cdot \mathrm{MeV}^{-3} \cdot (2J_r+1) \, 
\nonumber\\     
  &&\times \, M_T 
      \int_0^4 \xi d \xi \frac{1}{1 + \exp{\Delta(\xi - 1)}}\,
      K_1\left( {\textstyle{M_T\over T_{kin}}}\cosh\eta_t(\xi) \right)
\nonumber \\
  && \times \,
  I_0\left( {\textstyle{P_T\over T_{kin}}}\sinh\eta_t(\xi) \right) \, .
\end{eqnarray}
The integral over $\xi$ contains a Woods-Saxon spatial profile with a 
diffuseness parameter $\Delta$ and the two Bessel functions which originate 
from the boost of a thermal distribution with a transverse collective velocity.
The transverse expansion is described by a transverse rapidity, which is
parameterized as $\eta_t(\xi)=\eta_f \cdot \xi$, i.e. depending linearly 
on the normalized radial coordinate $\xi = r/R$. 
I fix $\Delta \equiv 50$ which yields a nearly
box-like spatial shape of the source, as this has been shown to be a 
reasonable assumption \cite{tomasik99,huovinen01,tp:hydro}.
A fixed upper limit ($\xi = 4$) has to be chosen for the numerical integration, 
for the distribution used in this work this is effectively equivalent to setting 
the upper limit to infinity.
A spin degeneracy factor is included, and there is an additional 
arbitrary normalization factor $N_{hydro}$, which is the same for all particle species
and controls the relative strength of the hydrodynamical component.
While the spectral shape is a.o. determined by the kinetic temperature
$T_{kin}$, the normalization for each particle species 
is readjusted to the chemical temperature $T_{chem}$ 
assuming that $dN/dy$ at 
midrapidity scales with the temperature as \cite{Schn93}:
\begin{equation} \label{dndythermal}
	{{dn_{\rm th}}\over{dy}} = { V \over {(2\pi)^2} } T^3
	\Big(
	{{ m^2 }\over{T^2}}  + {2m\over T}+ 2 
	\Big)
	\exp\left(-{m\over T} \right) \; .
\end{equation}
The parameterization attempts to describe the production of pions, protons and
antiprotons and kaons; it includes contributions 
from the following resonances: $\rho$, $\mathrm{K}^{0}_{S}$, 
$\mathrm{K}^{\star}$, $\Delta$, $\Sigma + \Lambda$, $\eta$, $\omega$, 
$\eta^{\prime}$. For the description of baryons a baryonic 
chemical potential $\mu_{B}$ is used, while for strange particles 
an additional strangeness suppression factor $\lambda_{s}$ is
introduced. For each hadron species the spectrum of hydrodynamically 
produced particles $f_{hydro}^{(X)}(p_{T})$ is given as the sum of the
direct contribution and those of the appropriate resonances:
\begin{eqnarray}
\label{eq:urs2}
  f_{hydro}^{(X)}(p_{T})
  &=& 
  {dN_X^{\rm dir}\over dy dm_T^2} + \,  \nonumber \\
  && + \sum_r \int_{\mathrm {phase} \, \mathrm{space}} dW^2 dY_r dM_{Tr} \cdot
  {F}(W,P_r,p)  \cdot {dN_r^{\rm dir}\over dY_r  dM_{Tr}^2},
\end{eqnarray}
where $W$ is the invariant mass, $Y_r$ the rapidity and $M_{Tr}$ the 
transverse mass of the resonance, and ${F}(W,P_r,p)$ is the appropriate 
phase space factor.

\subsection{The p+p parameterization}
\label{sec:pppar}

\begin{figure}[bt]
    \begin{center}
\includegraphics{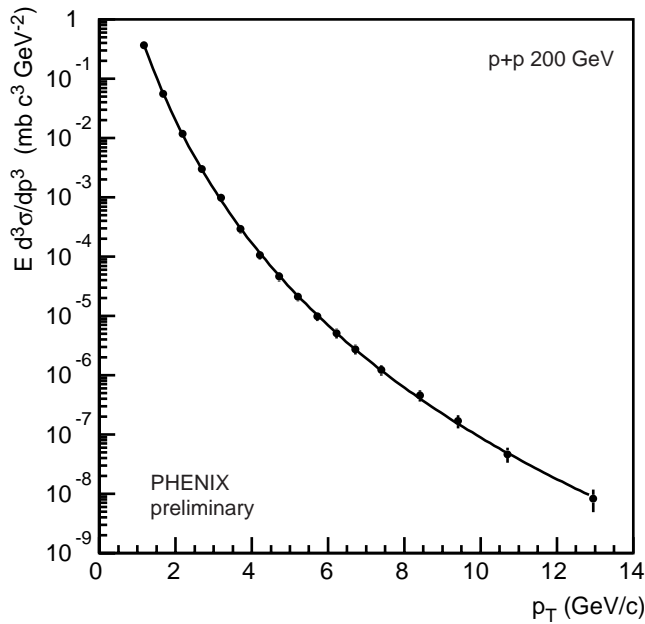}
\caption{Transverse momentum spectrum of neutral pions in p+p
collisions at $\sqrt{s} = 200 \, \mathrm{GeV}$ as measured by
the PHENIX experiment \protect\cite{qm02:hisa}. The solid
line shows a fit of equation~\protect\ref{eq:pppar1}.
}
	\protect\label{fig:ppfit}
    \end{center}
\end{figure}
As suggested by Hagedorn \cite{HAGEDORN83}, hadron spectra in high
energy p+p collisions can be described by a QCD-inspired power law with
exponential continuation at lower momenta. Such a spectral shape can 
be written as:  
\begin{equation}
    f(p_{T}) = 
    \left\{
        \begin{array}{l@{\quad:\quad}l}
            C \left( \frac{nT-p_{T1}}{nT} \right)^{n} \exp\left(
\frac{p_{T1}-p_{T}}{T} \right) & p_{T} \le p_{T1}  \\
            C \left( \frac{nT-p_{T1}}{nT-p_{T1}+p_{T}} \right)^{n} & 
p_{T} > p_{T1}.
        \end{array}
    \right.
    \label{eq:pppar1}
\end{equation}
Here $n$ and $T$ are adjustable parameters. $T$ is
related to the more frequently used parameter $p_{0}$ by:
$p_{0} = nT - p_{T1}$.
Fig.~\ref{fig:ppfit} shows a fit of this function to the neutral pion
spectrum measured in p+p collisions by the PHENIX experiment \cite{qm02:hisa}.
$p_{T1} = 1 \, \mathrm{GeV}/c$ was chosen. The function provides
a perfect fit over the full range of the measured data. As fit
parameters $C = 280.52 \, \mathrm{mb GeV^{-2}}$, $T = 0.226 \,
\mathrm{GeV}$ and $n = 9.89$ have been obtained. 

These distributions contain a component due to hard scattering processes, 
which is expected to lead to a power law shape dominating at high $p_{T}$.
At low $p_{T}$ production from string fragmentation, a soft process, will 
be much more abundant and therefore dominate there. 
For the later description of spectra in heavy-ion reactions I will 
attempt to separate the two components, although a clear separation 
directly from the spectra is not possible. However, 
the exponential shape at low $p_{T}$ is believed to be due to the soft 
component, while at very high $p_{T}$ the shape will turn into a pure
power law $\propto p_{T}^{-n}$. 
The separation is of course completely arbitrary when looking at the
fit to the data in Fig.~\ref{fig:ppfit} alone. In fact, there the 
exponential is not even fitted directly to the data and is only determined
as the differentiable continuation of the shape at higher $p_{T}$.
The value of the separation point $p_{T1} = 1 \, \mathrm{GeV}/c$ is 
not uniquely determined, but it yields a reasonable inverse slope parameter 
for the exponential.

The separation is necessary for a comparison to heavy ion collisions, 
because the different physical origin of the two components will
lead to different scaling behavior, as the number of hard collisions 
increases more rapidly with increasing mass of the nuclei than the number
of strings, i.e. the number of participant pairs.
I will assume the soft
contribution to be described by the exponential:
\begin{equation}
    s(p_{T}) = 
	    C_{s} \left( \frac{nT-p_{T1}}{nT} \right)^{n} \exp\left(
\frac{p_{T1}-p_{T}}{T} \right) 
    \label{eq:ppsoft}
\end{equation}
After subtraction of this part the hard component can be written as:
\begin{equation}
    h(p_{T}) = 
    \left\{
	\begin{array}{l@{\quad:\quad}l}
	    0 & p_{T} \le p_{T1}  \\
	    C_{h} \left[ \left( \frac{nT-p_{T1}}{nT-p_{T1}+p_{T}} \right)^{n}
	    - \left( \frac{nT-p_{T1}}{nT} \right)^{n} \exp\left(
	    \frac{p_{T1}-p_{T}}{T} \right) \right] & 
p_{T} > p_{T1}, 
	\end{array}
    \right.
    \label{eq:pphard}
\end{equation}
where $C_{s} = C_{h} = C$ for p+p collisions.

This parameterization with a subtracted exponential may look artificial. One 
should remember, however, that already the original \emph{power-law-like} shape
suggested by Hagedorn was an effective parameterization which turns into a 
power law for very high $p_{T}$. The deviation from the power law at lower 
momenta, also to be seen as a variation of the exponent with momentum, was 
not derived from underlying principles, but chosen to describe the experimental 
data. The hard component introduced here is equivalent at high $p_{T}$ and takes 
into account partial contributions of a soft component at intermediate $p_{T}$.

\begin{figure}[bt]
    \begin{center}
	\includegraphics{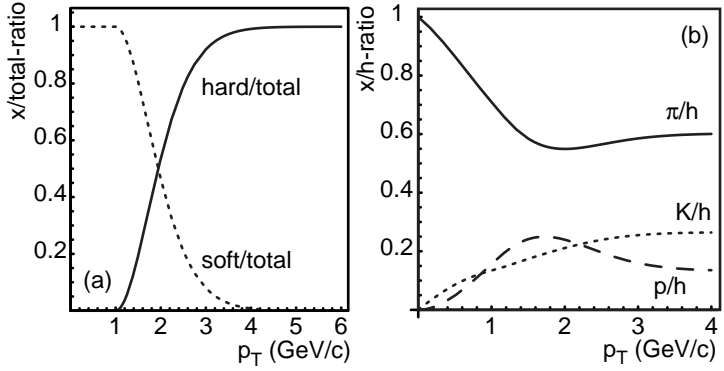}
	\caption{a) Assumed relative contribution of the soft and hard
	components of the pion spectrum in p+p.
	b) Fraction of the total number of charged hadrons belonging 
	to different hadron species. The solid line represents the
	pion fraction, the dotted line the kaons and the dashed line 
	the protons and antiprotons.}
	\protect\label{fig:pprat}
    \end{center}
\end{figure}

The pion spectrum is:
\begin{equation}
    f_{pp}^{(\pi)}(p_{T}) = s(p_{T}) + h(p_{T}).
    \label{eq:pispec}
\end{equation}
This is essentially only a rewriting of equation~\ref{eq:pppar1}.
The main purpose of this separation is the possibility to scale the
two components independently when increasing the system size by going
to heavy-ion collisions. In na\"{\i}ve pictures of particle
production the hard component is expected to increase with the number
of nucleon-nucleon collisions, as cross sections are small and processes are
incoherent. The soft component, however, will more likely scale with 
the number of participating nucleons as suggested by the wounded
nucleon model. The prescription used here leads to a continuous transition 
between the soft and hard component as can be seen in
Fig.~\ref{fig:pprat}a. The two components are of similar magnitude for 
$p_{T} \approx 2 \, \mathrm{GeV}/c$.

To describe the spectra of other hadron species further assumptions
are necessary. The only data from similar beam energies and
a reasonably large coverage in $p_{T}$ are available from the ISR
\cite{alper75}. One of the major observations in this paper was, that
the $p_{T}$-dependent ratios of different hadron species were to a good 
approximation independent of the beam energy studied. The kaon, proton and
antiproton spectra have thus been tuned to reproduce the observed
ratios. This has been achieved by using for kaons:
\begin{equation}
    f_{pp}^{(K)}(p_{T}) = 0.44 \cdot \tanh(0.46 \, \mathrm{GeV}^{-1}c p_{T}) \cdot s(p_{T}) + 
    0.44 \cdot h(p_{T}),
    \label{eq:kspec}
\end{equation}
for protons:
\begin{equation}
    f_{pp}^{(p)}(p_{T}) = \frac{1.8 \cdot \tanh(0.25 \, \mathrm{GeV}^{-2}c^{2} p_{T}^{2})}
    {1.3 + 0.4 \, \mathrm{GeV}^{-1}c p_{T}} \cdot s(p_{T}) + 
    0.33 \cdot h(p_{T}),
    \label{eq:pspec}
\end{equation}
and for antiprotons:
\begin{eqnarray}
    f_{pp}^{(\bar{p})}(p_{T}) &=& \frac{1.8 (0.3 + 0.4 \, \mathrm{GeV}^{-1}c p_{T}) 
    \cdot \tanh(0.25 \, \mathrm{GeV}^{-2}c^{2} p_{T}^{2})}
    {1.3 + 0.4 \, \mathrm{GeV}^{-1}c p_{T}} \cdot s(p_{T}) \nonumber\\
    && {} + 0.11 \cdot h(p_{T}).
    \label{eq:apspec}
\end{eqnarray}
The analytic expressions have been chosen because they can conveniently describe 
the shapes observed experimentally, they have no physical significance.
The parameters have been tuned to describe the experimentally
observed ratios obtained at the ISR (see Fig.~14 in \cite{alper75}).
They lead to particle ratios as shown in
Fig.~\ref{fig:pprat}b. No error analysis has been done for the parameters used in equations
\ref{eq:kspec}, \ref{eq:pspec} and \ref{eq:apspec}. Consequently the related
additional uncertainty in the final fits has not been taken into
account.

\section{Fits to experimental spectra}
\label{sec:fits}

\begin{table}[b]
\caption{Average number of NN collisions and participant nucleons 
for centrality classes as used by PHENIX \protect\cite{klaus}.}
\begin{tabular}[]{ccc}
    \hline
\hline
Centrality bin & $<N_{coll}>$ & $<N_{part}>$\\ \hline
0-5\%  &  1065.4 $\pm$ 105.3  &  351.4 $\pm$ 2.9 \\
5-10\%  &  845.4 $\pm$ 82.1  &  299.0 $\pm$ 3.8 \\
10-15\%  &  672.4 $\pm$ 66.8  &  253.9 $\pm$ 4.3 \\
15-20\%  &  532.7 $\pm$ 52.1  &  215.3 $\pm$ 5.3 \\
20-30\%  &  373.8 $\pm$ 39.6  &  166.6 $\pm$ 5.4 \\
30-40\%  &  219.8 $\pm$ 22.6  &  114.2 $\pm$ 4.4 \\
40-50\%  &  120.3 $\pm$ 13.7  &   74.4 $\pm$ 3.8 \\
50-60\%  &   61.0 $\pm$  9.9  &   45.5 $\pm$ 3.3 \\
60-70\%  &   28.5 $\pm$  7.6  &   25.7 $\pm$ 3.8 \\
70-80\%  &   12.4 $\pm$  4.2  &   13.4 $\pm$ 3.0 \\
\hline
\hline
\end{tabular}
\label{tab:Ncoll}
\end{table}

Fits of the function
\begin{equation}
    f^{(X)}(p_{T}) = f_{hydro}^{(X)} + f_{pp}^{(X)}
    \label{eq:fitfunc}
\end{equation}
have been performed to spectra of identified hadrons as measured by
the PHENIX experiment in Au+Au collisions at 200~GeV. Preliminary spectra of
charged pions, kaons, protons and antiprotons \cite{qm02:tatsuya} and
neutral pions \cite{qm02:david} have been 
used simultaneously for a given centrality. The centrality classes
used are summarized in Table~\ref{tab:Ncoll} together with estimates 
of the number of binary collisions $<N_{coll}>$ and the number of 
participants $<N_{part}>$ \cite{klaus}. Besides the parameters of the
hydrodynamical parameterization two normalization parameters $N_{s}
= C_{s}/C$ and $N_{h} = C_{h}/C$ of the pp parameterization enter.
All parameters are summarized in Table~\ref{tab:para}. Four
different sets of fits have been performed:
\begin{enumerate}
    \item  Pure hydrodynamic fits (i.e. $N_{s} = N_{h} = 0$) using only
    data for $m_{T} - m_{0} \le 1 \, \mathrm{GeV}/c^{2}$.

    \item  Pure hydrodynamics ($N_{s} = N_{h} = 0$) using the
    full data sets.

    \item  Pure fits of the pp parameterization ($N_{hydro} = 0$)
    using the full data sets.

    \item  Fits of the full hybrid parameterization 
    using the full data sets.

\end{enumerate}
The hydrodynamic fits (1) to the low momentum region provide
excellent fits for all centralities. There is some ambiguity in the choice
of $T_{kin}$ and $\langle \beta_T \rangle$ -- to some extent 
a large velocity can be
compensated by a small temperature, and vice versa. To avoid these
ambiguities all later fits have been performed with setting 
$T_{kin} \equiv 120 \, \mathrm{MeV}$, which provides good agreement at low
$m_{T}$ for all centralities. One should note that while especially
the flow velocity depends on the choice of the freeze-out
temperature, the results on the scaling of the different particle
species as discussed below are not strongly affected. 

\begin{table}[h]
\caption{Fit parameters of the proposed parameterization. Note, that
while the average velocity is given here, this has been calculated from the
transverse rapidity parameter $\eta_{f}$ which was used in the fits.
Fit parameter values for central collisions with their fit errors are
stated in the right column.}
\begin{tabular}[]{llc}
\hline
\hline
parameter & remark & fit results (central) \\ \hline
$N_{hydro}$  & normalization of hydrodynamical contribution & $7.7 \pm 3.1$    \\
$T_{kin}$  & kinetic temperature & $120 \, \mathrm{MeV}$ \\
$\langle \beta_T \rangle$  & average transverse expansion velocity
  & $0.529 \pm 0.015$ \\
$T_{chem}$  & chemical temperature & $160.8 \pm 4.3 \, \mathrm{MeV}$  \\
$\mu_{B}/T_{chem}$  & baryonic chemical potential & $0.17 \pm 0.03$ \\
$\lambda_{s}$  & strangeness suppression & $0.96 \pm 0.15$ \\ \hline
$N_{s}$  & normalization of soft component & $124 \pm 100$  \\
$N_{h}$  & normalization of hard component & $173 \pm 20$ \\
\hline
\hline
\end{tabular}
\label{tab:para}
\end{table}

\begin{figure}[bt]
    \begin{center}
	\includegraphics{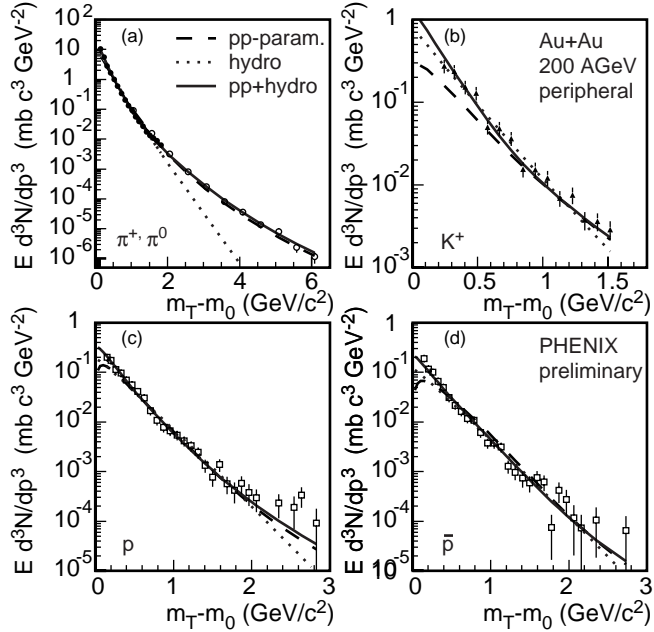}
	\caption{Transverse momentum spectrum of pions, kaons,
	protons and antiprotons in peripheral Au+Au
	collisions ($70-80 \% $) at $\sqrt{s} = 200 \, \mathrm{GeV}$ 
	as measured by
	the PHENIX experiment \protect\cite{qm02:tatsuya,qm02:david}. 
	The dotted line shows pure hydrodynamics fits (2), the dashed
	line a fit of the pure pp parameterization (3) and the solid
	line shows a full fit (4) of equation~\protect\ref{eq:fitfunc}.
	}
	\protect\label{fig:specper}
    \end{center}
\end{figure}

Examples of these fits to the spectra from peripheral collisions ($70-80 \% $)  
are shown in
Fig.~\ref{fig:specper}. The pion spectra can be well described
already by the pure pp parameterization as well as the full fit, the 
pure hydrodynamics fit, however, fails to describe the high $m_{T}$ tail.
While for pions all parameterizations provide a good description at
lower $m_{T}$, for kaons, proton and antiprotons discrepancies
are seen mostly at low momenta. These discrepancies are largest for
the pp parameterization. As I am most interested in the behavior at 
high momenta, this will not be investigated further. One may 
take it as a hint that low momentum particle ratios extracted from pp
measurements at ISR \cite{alper75} do not completely describe even
peripheral heavy ion collisions at RHIC, while the spectral shape of 
pions appears to be well described over the range investigated here. 
The reasonable description of all particle species at intermediate
$m_{T}$ lends some support to the choice of the pp parameterization.
The full fit combining hydrodynamics and the pp parameterization
provides an excellent description of all spectra.
Of course, the high momentum behavior of the heavier particles can
not be thoroughly tested from the limited momentum range alone.

\begin{figure}[bt]
    \begin{center}
	\includegraphics{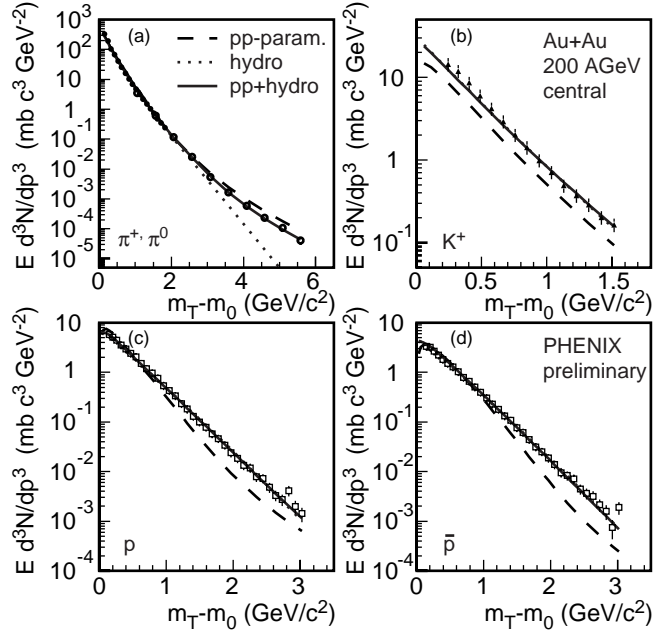}
	\caption{Transverse momentum spectrum of pions, kaons,
	protons and antiprotons in central Au+Au
	collisions ($0-5 \% $)  at $\sqrt{s} = 200 \, \mathrm{GeV}$ 
	as measured by
	the PHENIX experiment \protect\cite{qm02:tatsuya,qm02:david}. 
	The dotted line shows pure hydrodynamics fits (2), the dashed
	line a fit of the pure pp parameterization (3) and the solid
	line shows a full fit (4) of equation~\protect\ref{eq:fitfunc}.
	}
	\protect\label{fig:speccen}
    \end{center}
\end{figure}

Similarly, Fig.~\ref{fig:speccen} shows fits to spectra for central
collisions ($0-5 \% $). Again, it can be seen that hydrodynamics alone fails to
describe the tail of the pion spectra, while the other parameterizations
do a reasonably good job. However, for the heavier particles the
failure of the pure pp parameterization is obvious. There are large
discrepancies for the kaons over the whole momentum range and for
protons and antiprotons for large momenta. It is clear that the
spectra for central collisions can not be described neither by hydrodynamics 
nor a pure pp parameterization alone, however the combination of both 
provides again an excellent description. An equally good description 
can be obtained with this parameterization for all centralities. 

\section{Discussion}
\label{sec:discuss}

\begin{figure}[bt]
    \begin{center}
	\includegraphics{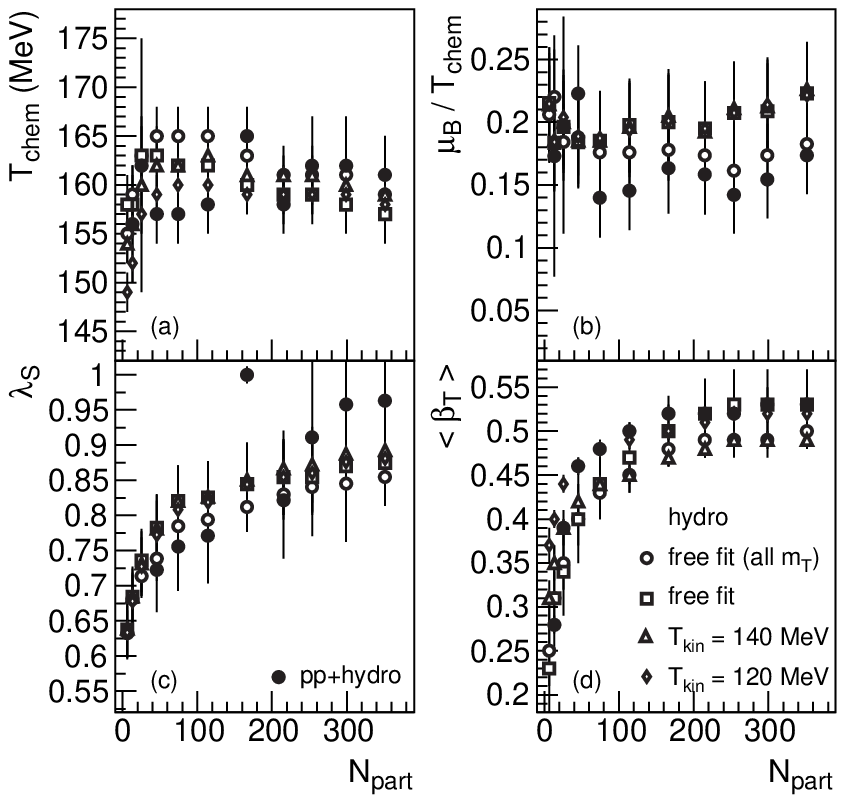}
	\caption{Parameters of the hydrodynamical component of fits
	to Au+Au
	collisions at $\sqrt{s} = 200 \, \mathrm{GeV}$ as a function
	of the number of participants. 
	a) chemical temperature, b) baryonic chemical potential
	normalized to the chemical temperature, c) strangeness
	suppression factor, d) average transverse expansion velocity.
	The open symbols show the results from pure hydrodynamic
	fits under different conditions, the filled circles show
	results using the full parameterization.
	}
	\protect\label{fig:thermpar}
    \end{center}
\end{figure}

\begin{figure}[t]
    \begin{center}
	\includegraphics{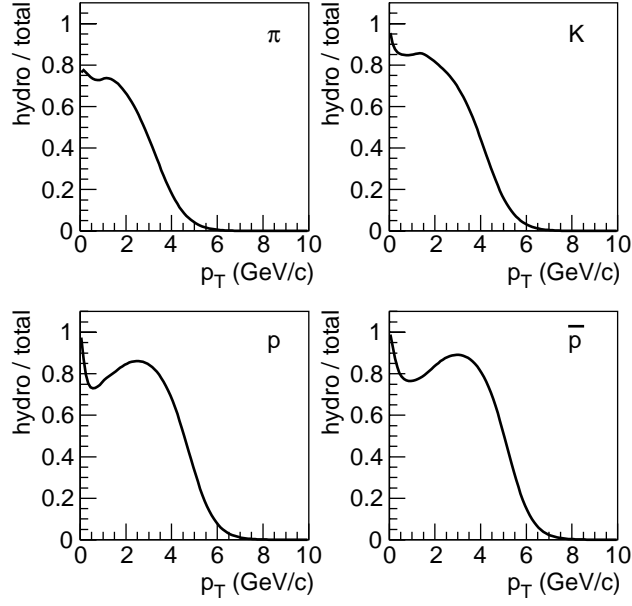}
	\caption{Ratio of the hydrodynamical contribution to the
	total spectrum in central Au+Au
	collisions at $\sqrt{s} = 200 \, \mathrm{GeV}$ as a function
	of $p_{T}$. 
	The ratios for pions, kaons, protons and antiprotons are
	shown separately.
	}
	\protect\label{fig:thermrat}
    \end{center}
\end{figure}

\begin{figure}[b]
    \begin{center}
	\includegraphics{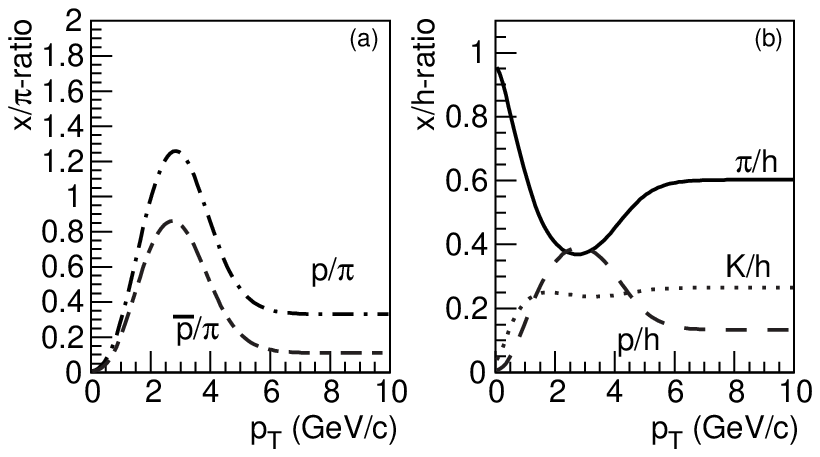}
	\caption{Hadron ratios in central Au+Au
	collisions at $\sqrt{s} = 200 \, \mathrm{GeV}$ as a function
	of $p_{T}$. 
	a) Protons/pions and antiprotons/pions, b) charged
	pions/charged hadrons, charged kaons/charged hadrons and
	proton+antiprotons/charged hadrons.
	}
	\protect\label{fig:auaurat}
    \end{center}
\end{figure}

Although the description of spectra by means of hydrodynamic
parameterizations is not the ultimate aim of this paper, it is still 
instructive to study the behavior of the fit parameters for the
different centralities. Fig.~\ref{fig:thermpar} shows parameter values 
for different fits as a function of centrality. The open symbols
show results from pure hydrodynamic fits. Those have been performed
for lower $m_{T}$ only 
with the kinetic temperature being either fixed to $T_{kin} = 120$ or
$140$~MeV or varying freely, in addition a fit with unconstrained
$T_{kin}$ has been performed to the full available $m_{T}$ range. The
filled circles show the same parameters for a full fit of 
equation~\ref{eq:fitfunc} including the soft and hard components and assuming 
$T_{kin} = 120 \, \mathrm{MeV}$. It 
is noteworthy that all these fits show essentially very similar results.
The chemical temperature (Fig.~\ref{fig:thermpar}a) is $T_{chem} \approx 160 \, \mathrm{MeV}$
for all centralities, also the baryonic chemical potential (b) appears to
be almost independent of centrality with a value of 
$\mu_{B}/T_{chem} \approx 0.15 - 0.2$. The strangeness suppression
(c) is
strongest for peripheral collisions with $\lambda_{s} \approx 0.65$
increasing to $\lambda_{s} \approx 0.9$ for central collisions. 
The transverse flow velocity also increases considerably with
centrality from $\langle \beta_T \rangle \approx 0.25$ for peripheral
reactions to $\langle \beta_T \rangle \approx 0.5$ in central
reactions. The chemical temperature appears to be similar, 
but slightly smaller than values
obtained for central collisions of Au+Au at 130~GeV \cite{tp:hydro,pbm-rhic}.
Kinetic temperature and flow velocity are similar to values obtained 
in \cite{tp:hydro} for 130~GeV or in \cite{qm02:jane} for 200~GeV,
when considering that the latter analysis did not take the influence 
of resonances into account. The results of the hydrodynamical
parameters obtained here are therefore not at all extraordinary or
astonishing. (For reference the fit values for central collisions using the 
full fit are also given in Table~\ref{tab:para}.) 

\begin{figure}[bt]
    \begin{center}
	\includegraphics{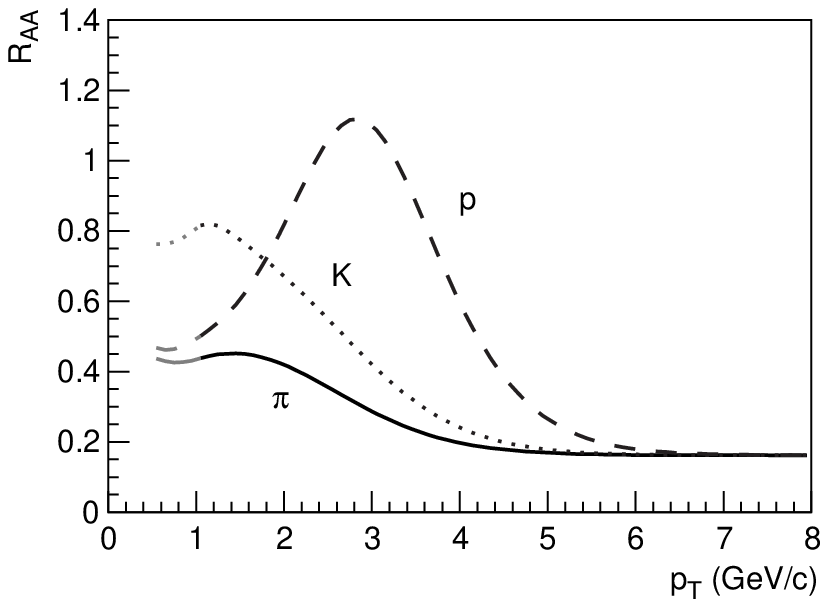}
	\caption{Nuclear modification factor (equation \ref{eq:raa}) as a function
	of $p_{T}$ for pions, kaons and protons+antiprotons.
	}
	\protect\label{fig:raa}
    \end{center}
\end{figure}

Already in \cite{tp:hydro} it was argued that the hydrodynamic
contribution to the hadron yield even beyond $p_{T} = 3 \,
\mathrm{GeV}/c$ is not negligible. This question can be revisited
with the help of the refined fits performed here. Fig.~\ref{fig:thermrat}
shows the relative contributions of the hydrodynamical component to
the full fits for central collisions. Again, it can be seen that this
contribution is significant for pions beyond $p_{T} = 3 \, \mathrm{GeV}/c$.
Moreover, the importance of this contributions reaches out to even
higher $p_{T}$ for the heavier particles, as is expected for a
hydrodynamic source. \footnote{The strong increase of the relative importance of
the hydrodynamic contribution in protons and antiprotons at low $p_{T}$ 
is due to the suppression of these particles in the non-hydro production 
(see Fig.~\protect\ref{fig:pprat}).}
This hydrodynamic component causes very naturally a different
behavior of the particle ratios compared to p+p collisions. The
proton/pion and antiproton/pion ratios (as displayed in
Fig.~\ref{fig:auaurat}a) show pronounced maxima close
to a value of 1 between $p_{T}$ of 2 and 4 GeV$/c$. Consequently there is a 
strong minimum in the pion/hadron ratio (see
Fig.~\ref{fig:auaurat}b).
The kaon/hadron ratio rises much faster than in p+p.

This will also influence the comparison of spectra from
central nuclear collisions to those from p+p collisions as is
customary in the investigations of high $p_{T}$ hadron suppression.
One uses the \textit{nuclear modification factor}:
\begin{equation} 
R_{AA}^{X}(p_T) = 
\frac{(1/N^{evt}_{AA})\,d^2N^{X}_{AA}/dy dp_T}{\langle N_{coll}\rangle \, 
(1/N^{evt}_{pp})\,d^2N^{X}_{pp}/dy dp_T}.
\label{eq:raa}
\end{equation}
This can easily be calculated from the parameterization used here.
Fig.~\ref{fig:raa} shows $R_{AA}$ for the different hadron species.
$R_{AA}$ for pions decreases from about 0.4 at 2~GeV$/c$ to 0.2 at
high $p_{T}$. For heavier particles $R_{AA}$ starts out at higher
values and reaches the same asymptotic
value more slowly. For protons and antiprotons there is a peak 
structure with a small enhancement ($R_{AA} > 1$) in the intermediate
$p_{T}$ range. 

\begin{figure}[bt]
    \begin{center}
	\includegraphics{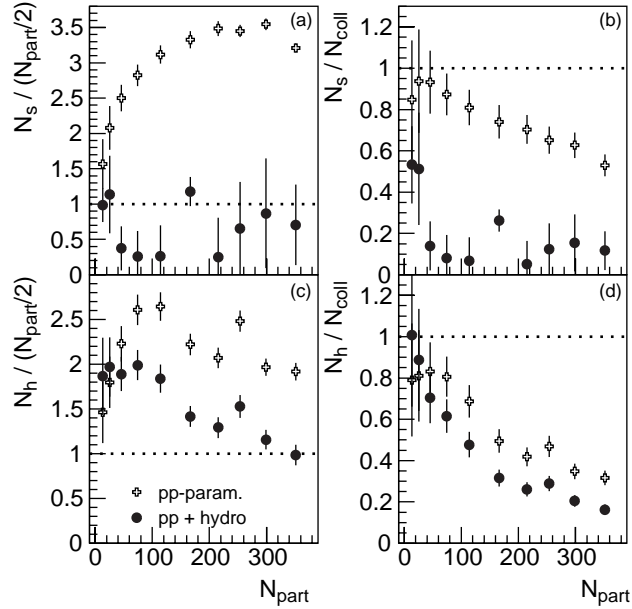}
	\caption{Strength of the soft and hard component as a
	function of the number of participants. The open crosses show
	the pure pp parameterization, the filled circles the full fit
	including the pp parameterization and hydrodynamic contributions. 
	a) $N_{s}/(0.5*N_{part})$, b) $N_{s}/N_{coll}$,
	c) $N_{h}/(0.5*N_{part})$, d) $N_{h}/N_{coll}$}.
	\protect\label{fig:hardsoft}
    \end{center}
\end{figure}

Finally one can investigate the relative strength of the soft and
especially the hard component, $N_{s}$ and $N_{h}$ as a function of
centrality. Fig.~\ref{fig:hardsoft}a shows $N_{s}/(0.5 \times N_{part})$ as a
function of $N_{part}$. Without taking into account any 
hydrodynamic production this normalized strength of the soft contribution
increases significantly with increasing centrality, while for the
full parameterization the ratio stays close or below a value of 1.
The difference between the two estimates may just be seen as the
importance of the added hydrodynamic contribution, which very
naturally appears to be negligible for the most peripheral collisions
and increases strongly with centrality. A similar difference is seen 
when normalizing to $N_{coll}$ (Fig.~\ref{fig:hardsoft}b), however,
now all values are smaller than 1. Even more interesting is the
evolution of the hard component. Fig.~\ref{fig:hardsoft}d shows 
$N_{h}/N_{coll}$ as a function of $N_{part}$. This can effectively 
be seen as an average value of $R_{AA}$ over the $p_{T}$ range of the hard
component. Qualitatively the trend of both parameterizations is
similar, there are, however, significant quantitative differences.
For the pure pp parameterization the apparent average suppression changes
from 0.8 for peripheral to 0.4 for central reactions. When
accounting for the hydrodynamic contribution, the ratio changes
more drastically from no suppression ($=1$) to a value of 0.2. Within
this model, these are the true suppression values of the
hard component independent of species and $p_{T}$.

\begin{figure}[tb]
    \begin{center}
	\includegraphics{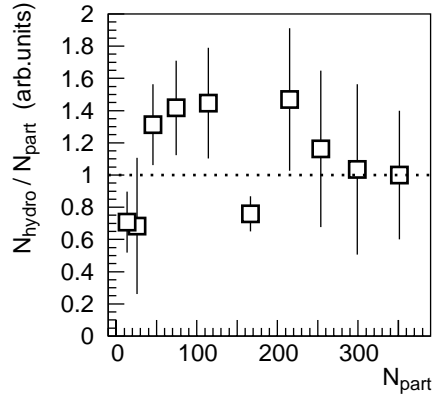}
	\caption{Strength of the hydrodynamic component per participant as a
	function of the number of participants. The ratio is normalized to one 
	for the most central collisions (see Table~\protect\ref{tab:para}.}
	\protect\label{fig:hydronorm}
    \end{center}
\end{figure}

The normalizations of the soft and hard component are well defined 
from their fit to pp data, the hydrodynamic component has an overall 
arbitrary normalization related e.g. to the unkown source volume. Once the 
normalization is fixed for a given centrality, it is still instructive to 
compare it to those for other centralities. Fig.~\ref{fig:hydronorm} shows 
$N_{hydro}/N_{part}$ normalized to one for the most central collisions
as a function of the the number of participants. While $N_{part}$ varies
by a factor of almost 30 over the centrality range investigated, the ratio 
appears to be constant within errors, i.e. the size of the hydrodynamic 
system is to a large extent determined from the collision geometry. This
is another hint for the consistency of the model used. 
Within this parameterization one can deal simultaneously with a 
suppression of the hard scattering component and a build up of 
hydrodynamic behavior. It would of course be of great interest to study 
explicitly the 
dynamics of the interactions between jets and the hydrodynamic system as
e.g. performed in \cite{Hir02} - this is however far beyond the 
scope of this paper.

\section{Summary}
Fits of a hybrid parameterization containing both hydrodynamics and
elementary soft and hard contributions to hadron spectra measured in 
Au+Au collisions at 
$\sqrt{s_{_{NN}}}$ = 200~GeV~\cite{qm02:tatsuya,qm02:david} 
have been performed.
The input distributions have been tuned to describe the spectra of
neutral pions in p+p collisions at $\sqrt{s}$ = 200~GeV~\cite{qm02:hisa} 
and the particle ratios as measured at the ISR~\cite{alper75}.
Keeping the parameters of the hydrodynamic source and the relative
normalizations of the soft and hard components as free parameters
an excellent description of hadron spectra for all centralities can 
be obtained. The parameters obtained for the hydrodynamic source are 
reasonable: The chemical temperature of $T_{chem} \approx 160 \, \mathrm{MeV}$
and the baryonic chemical potential of 
$\mu_{B}/T_{chem} \approx 0.15 - 0.2$ 
are almost independent of centrality. For peripheral collisions the 
strangeness suppression factor is $\lambda_{s} \approx 0.65$ and the 
flow velocity is $\langle \beta_T \rangle \approx 0.25$, while for
central collisions I obtain $\lambda_{s} \approx 0.9$ and
$\langle \beta_T \rangle \approx 0.5$. 

The hydrodynamic source contributes a significant fraction of
hadrons at intermediate $p_{T}$ reaching out to at least $p_{T} = 3
\, \mathrm{GeV}/c$ for pions and $p_{T} = 5 \, \mathrm{GeV}/c$ for
protons and antiprotons, thereby explaining 
the large baryon-to-pion ratio in central Au+Au collisions.
It also results in a very different scaling behavior for the
different species, even if one assumes one universal suppression
factor of the hard component for all particle species. The
suppression would only be visible for proton and antiprotons for
$p_{T} > 4 \, \mathrm{GeV}/c$. The universal suppression factor
extracted is smaller than 0.2 for central collisions, the merging of
$R_{AA}$ for different species can be seen for 
$p_{T} > 6 \, \mathrm{GeV}/c$.

\end{document}